# SCIENTIFIC REPORTS

**OPEN**

# Improved Self-cleaning Properties of an Efficient and Easy to Scale up TiO$_2$ Thin Films Prepared by Adsorptive Self-Assembly

Rima J. Isaifan[1,2], Ayman Samara[1], Wafa Suwaileh[1], Daniel Johnson[1], Wubulikasimu Yiming[3], Amir A. Abdallah[1] & Brahim Aïssa[1,2]

Transparent titania coatings have self-cleaning and anti-reflection properties (AR) that are of great importance to minimize soiling effect on photovoltaic modules. In this work, TiO$_2$ nanocolloids prepared by polyol reduction method were successfully used as coating thin films onto borosilicate glass substrates via adsorptive self-assembly process. The nanocolloids were characterized by transmission electron microscopy and x-ray diffraction. The average particle size was around 2.6 nm. The films which have an average thickness of 76.2 nm and refractive index of 1.51 showed distinctive anti soiling properties under desert environment. The film surface topography, uniformity, wettability, thickness and refractive index were characterized using x-ray diffraction, atomic force microscopy, scanning electron microscopy, water contact angle measurements and ellipsometry. The self-cleaning properties were investigated by optical microscopy and UV-Vis spectroscopy. The optical images show 56% reduction of dust deposition rate over the coated surfaces compared with bare glass substrates after 7 days of soiling. The transmission optical spectra of these films collected at normal incidence angle show high anti-reflection properties with the coated substrates having transmission loss of less than 6% compared to bare clean glass.

Sunlight is an abundant energy source that is not distributed evenly on earth's surface. Low latitudes, arid and semi-arid areas, within 35°N to 35°S, receive the highest direct normal irradiance[1]. Seven of the world deserts, located between these two latitudes, are able to meet the energy needs globally with solar power generation technologies, including photovoltaic (PV), concentrated photovoltaic (CPV), and concentrated solar power (CSP) systems[1]. Nevertheless, solar cells and solar collectors in sandy areas, e.g. the Middle East and North Africa (MENA) regions are subjected to severe deposition of sand and dust particles on the glass surface of the cells or collectors in a process called fouling[2]. Fouling is defined as the deposition of foreign particles such as fine sand particles on a heat transfer surface forming an insulating layer that reduces the rate of heat transfer[2,3].

Although several ways have been reported so far to clean soiled surfaces such as the use of electrostatic dust removal[4], mechanical removal methods that are mainly comprised of a broom or a brush driven by a machine[5], and the use of surfactants[2], it should be noted that solar panels are usually mounted in regions with difficult accessibility such as the roof of buildings, making thereby the solution based on self-cleaning coatings highly preferable compared to conventional cleaning methods.

Self-cleaning nanofilms were developed to be coated on solar cell array. These films are made of hydrophobic or hydrophilic material which involves two different strategies. In the case of hydrophobic films, studies have suggested that super hydrophobic surfaces form microstructures and/or nanostructures that enhance the contact angle of water. So, when water droplets hit the surface, it would quickly roll off, carrying dust and other particles away[6].

On the other hand, the most popular hydrophilic material studied for self-cleaning applications is titanium dioxide (TiO$_2$)[7–10]. Self-cleaning using TiO$_2$ films involves two stages; first, a split of organic dirt via photocatalytic process in the presence of ultraviolet light, and second, the diffusal of water to the whole surface instead of getting together due to the hydrophilicity of TiO$_2$ and hence rinsing the dust. In addition to its hydrophilic properties and high photo reactivity, titania also exhibits long term stability, good mechanical and chemical properties, in addition to high thermal resistance together with low toxicity and costs[11].

[1]Qatar Environment and Energy Research Institute, Hamad Bin Khalifa University, Qatar Foundation, P.O. Box, 5825, Doha, Qatar. [2]College of Science and Engineering, Hamad Bin Khalifa University, Qatar Foundation, P.O. Box, 5825, Doha, Qatar. [3]Office of Research, Texas A & M University at Qatar, Doha, Qatar. Correspondence and requests for materials should be addressed to R.J.I. (email: risaifan@hbku.edu.qa)

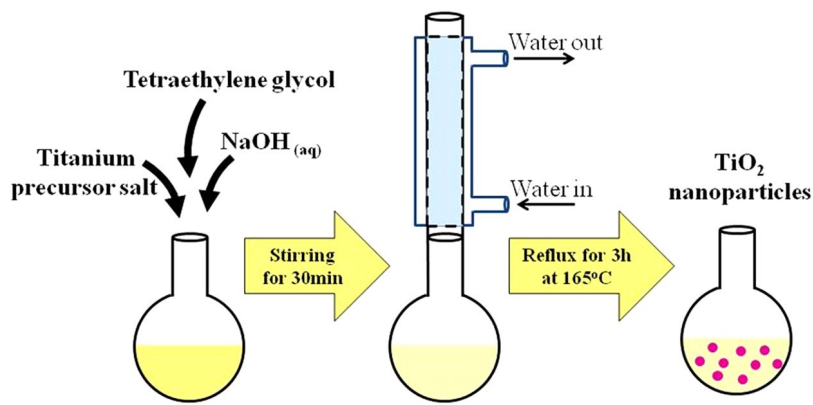

**Figure 1.** Illustration of TiO$_2$ nanoparticle synthesis via polyol reduction method.

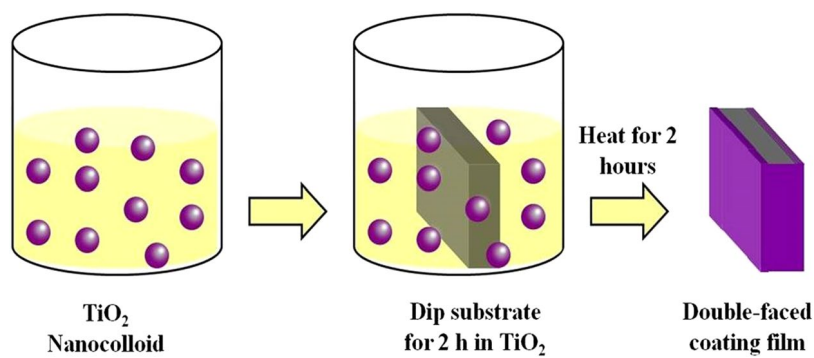

**Figure 2.** Illustration of coating preparation steps via adsorptive self-assembly method.

The functional characteristics of titania is related to its crystal structure and morphology which both depend on the preparation method. In order to reduce the accumulation of contaminants on PV surfaces, self-cleaning properties should be combined with the efficient photo catalytic activity of titania. Both self-cleaning and AR properties depend on the surface nano-texture and roughness[10]. The latter influences the optical properties of the thin film and depends on the preparation procedure as well[12]. Since this application would be of interest for industrial scalability of the procedure as well as cost efficiency, focus was put to utilize a simple, controllable, cost effective and an easy to scale-up coating method; that is the adsorptive self-assembly with thermal annealing of the film which was first developed by Xi *et al.*[8]. This low temperature synthesis method is also suitable to prepare many other coating layers.

In this work, TiO$_2$ nanocolloids were prepared via modified polyol reduction method and used as self cleaning coating for glass substrates. These coatings were characterized by atomic x-ray diffraction, force microscopy, scanning electron microscopy, ellipsometry, optical microscope, UV-Vis spectroscopy and contact angle measurements. Moreover, the effect of soiling on the optical transmission properties of the coated glass was investigated and the effectiveness of the anti- soiling properties was evaluated and compared to uncoated glass surfaces.

## Experimental

**TiO$_2$ nanocolloids synthesis.** TiO$_2$ nanocolloids were prepared by polyol reduction method[13–15]. First, 125 mL of tetraethyleneglycol (TEG) (Sigma-Aldrich, ≥99%) were measured and poured in a three-neck flask. Then, 2.7 g of titanium (IV) oxysulfate TiOSO$_4$ (Sigma-Aldrich 99%) precursor salt were measured and introduced into the three-neck flask and stirred for 30 min to dissolve in the TEG at room temperature. TEG is used as a solvent and as a reducing agent to the precursor salt[15, 16]. To control the particle size of the synthesized nanoparticles, 2.9 g of sodium hydroxide pellets were dissolved in 5 mL of deionized water and the solution was gradually added to the mixture using a syringe. The mixture was mechanically stirred and heated at a rate of 6 °C min$^{-1}$ from room temperature to 165 °C under reflux for 3 hours. At this stage, TiO$_2$ nanocollids were synthesized (Fig. 1). These colloids are used as is for the coating of substrates as per Fig. 2.

In order to obtain powder nanoparticles for x-ray diffraction analysis, the prepared colloidal suspension containing the residual polyol and the nanoparticles were left to cool down at room temperature, and then centrifuged at a speed of 5000 rpm for 20 min followed by washing with pure ethanol. The procedure of centrifugation and ethanol-washing was repeated five times. Finally, the obtained powders were dried in an oven (Thermolyne, Thermo Scientific) overnight at 50 °C.

**Preparation of coating films.** Borosilicate plate glass slides (Chemglass CG-1904-36) of 25 mm × 10 mm × 2 mm (width × length × thickness) were used as work pieces. Before deposition, the glass substrate samples were cleaned with ethanol and rinsed with deionized water. Then, each substrate sample was soaked in Piranha solution which was prepared by mixing sulfuric acid ($H_2SO_4$, concentration of 5–6%, Merck) with hydrogen peroxide ($H_2O_2$, concentration of 30%, Merck) in a volume ratio of 2:1 for 30 minutes. Later, the samples were thoroughly rinsed with deionized water and left to dry in the oven at 70 °C for another 30 minutes. The dried glasses were then dipped in nanocolloids for 2 hours (Fig. 2) at 20 °C and at a relative humidity of around 30%. The samples were then placed in an oven (Thermolyne, Thermo Scientific) at a temperature of 400 °C under air for 2 hours following the procedure depicted in reference[8]. The prepared sample had a replicate of 3 to assure reproducibility of the experimental results.

**Characterization techniques.** The transmission electron microscopy (TEM) analysis was carried out using FEI TALOS X operated at 200 kV. The TEM specimens were prepared by sonicating the as-prepared $TiO_2$ catalyst powder in ethanol. One drop of the solution was then placed onto a 200 mesh TEM copper grid coated with a lacey carbon support film (Ted Pella) and dried in air. We used Image J software for the determination of particles size distribution.

X-ray diffraction (XRD) patterns of the synthesized nanoparticles and thin films were acquired using a Rigaku Ultima IV X-ray diffractometer equipped with fixed monochromator which was used for the data collection. The XRD was operated at 40 KV and 40 mA with divergence slit and scattering slit of 2/3degree, divergence height limiting slit of 10 mm and a receiving slit of 0.3 mm. Continuous scans on the samples were carried out between 10–90 2θ degree with 0.02 degree step size and 1 degree/minute data collection time. The crystallite size of the $TiO_2$ nanoparticles was estimated from XRD line broadening using Scherrer equation[17].

$$D = \frac{K\lambda}{FWHM \cos\theta} \quad (1)$$

where, D is the averaged dimension of crystallites; K is the Scherrer constant, an arbitrary value that falls in the range 0.87–1.0 (usually assumed to be 0.9); λ is the wavelength of X-ray; and FWHM is pure diffraction broadening of a peak at half height, due to crystallite dimensions located at 2θ.

Scanning Electron Microscopy (SEM) images were obtained using SEM Model JCM-6000PLUS NeoScope to observe the morphology of coating layers of $TiO_2$ in different positions of the glass cross sections. SEM/SEI topographical images were obtained on the three glass samples with magnification of 500X and 200X and resolution of 256 × 192 Pixels at an accelerating voltage 10 kV, an energy range 0–20 KeV and high vacuum mode.

AFM characterization was carried out using a Dimension Icon model AFM with NanoScope V Controller (Bruker AXS, USA) operating in PeakForce mode. All measurements were made in ambient conditions using NSG30 silicon tapping mode probes (NT-MDT, Russia).

Optical transmittance spectra of the coated samples were recorded in the 200–1000 nm range using a Jenway-67 Series Spectrophotometer. Optical analysis of the coated surfaces was performed using Olympus (IX73) optical microscope. The objective lens of 40X magnification was used. The surface wettability was evaluated by measuring the contact angle of deionized water droplets deposited on the film surface under ambient conditions using Rame-Hart with three replicates. The acquired images have been elaborated with Drop Image software to obtain the average contact angles. The thin film thickness and refractive index for all samples were measured using J.A. Woollam WVASE Ellipsometer. Each sample was tested using 2 ellipsometer angles (65° and 70°), and light wavelength scan between 300 and 1000 nm. The ellipsometer measures the changes in the light polarization state and generates a graph of the changes in ellipsometeric angles (Δ, ψ) as a function of wavelength. In the analyses of the ellipsometric data, the samples were treated as composed of a $TiO_2$ thin film on a thick $SiO_2$ substrate. For simplicity, a basic Cauchy model for transparent films was used to fit the data, and the average thickness and refractive index of the $TiO_2$ layer were calculated numerically from the ψ and Δ functions[18].

## Results and Discussion

Figure 3 shows bright field TEM and high resolution TEM images of the $TiO_2$ colloidal particles where the black localized features corresponding to the $TiO_2$ nanoparticles. Typically, the nanopartciles sizes were found to be uniform and spherical in shape, with the presence of few non-uniform particles resulting most probably from agglomeration process. The size of the nano-particles ranges from 2 to 5 nm with an average particle size estimated to be around 2.61 nm. This average size value is well corroborated by the crystal size as estimated by XRD analysis and shown in Table 1.

The X-ray diffraction patterns of the synthesized titania nanoparticles and the deposited $TiO_2$ thin film are shown in Fig. 4a and b, respectively. The peak details are summarized in Table 1. The experimental XRD pattern is in agreement with the JCPDS card no. 21–1272 (anatase $TiO_2$) and the XRD pattern of $TiO_2$ nanoparticles as reported in literature[19, 20]. The 2θ peaks located at around 25° and 48° confirm the $TiO_2$ anatase structure[19, 20]. The high intensity of XRD peaks reflects that the formed nanoparticles are crystalline and broad diffraction peaks indicate very small size crystallite.

Figure 5 shows a top view AFM image of the $TiO_2$ thin films on glass substrate. The image shows dense and compact surface morphology. Moreover, the surface roughness was obtained of the coated surfaced to be 4.32 nm$^2$.

Figure 6(a) and (b) shows representative SEM images of the uncoated and coated samples after soiling (dust accumulation) for 7 days, respectively. It is clearly seen that the dust particles deposited on the uncoated sample surface (Fig. 6a) are dense and agglomerated while those settled on the coated glass sample surface (Fig. 6b) are scares and dispersed. Figure 6c and d show optical images at 40X magnification object lens of the uncoated and coated samples, respectively. The titania coating deposited using TiO$_2$ colloidal nanoparticles shows less accumulation of dust particles compared with uncoated glass sample.

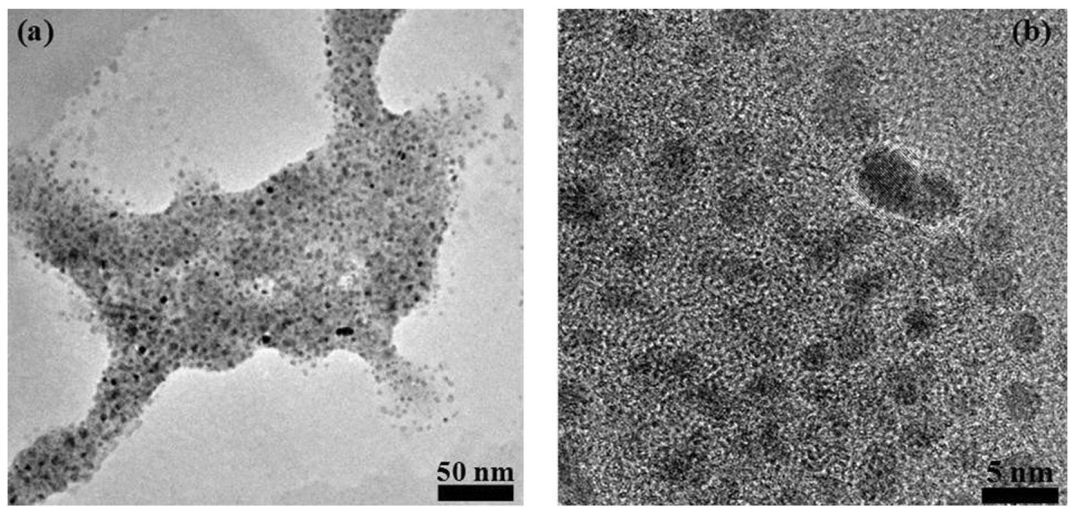

**Figure 3.** (**a**) TEM and (**b**) HRTEM micrographs of the colloidal TiO$_2$ nanoparticles.

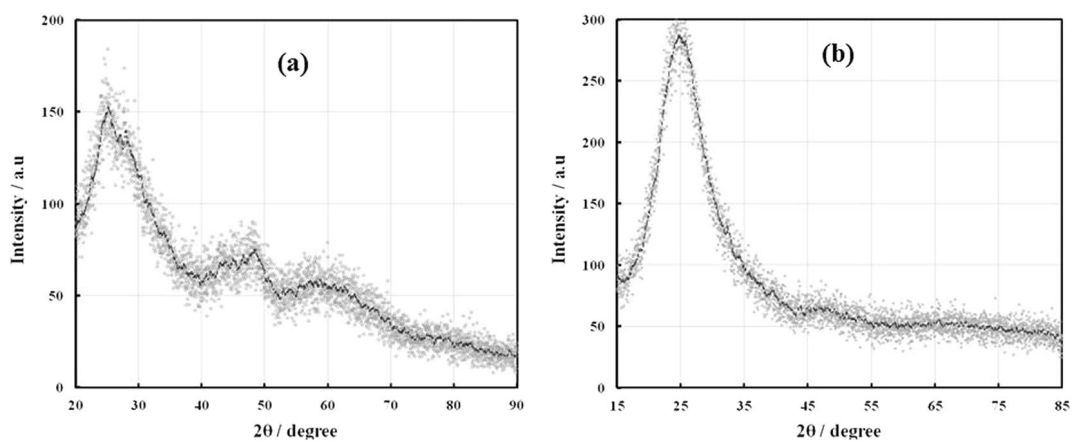

**Figure 4.** X-ray diffraction pattern of (**a**) the synthesized TiO$_2$ nanoparticles and (**b**) the deposited TiO$_2$ thin film.

| Nanoparticle | 2 θ max (°) | FWHM (°) | Crystal size (nm) |
|---|---|---|---|
| TiO$_2$ | 25.0 | 7.5 | 1.13 |
|  | 48. 2 | 3.3 | 2.76 |

**Table 1.** XRD characteristics of Pt colloids by polyol reduction method.

Moreover, Fig. 7 shows the rate of dust deposition per surface area of glass substrate after the samples were left in the field for 7 days soiling. Dust deposition rate was obtained through the following equation:

$$Dust\ deposition\ rate\left(\frac{g}{cm^2}\right) = \frac{soiled\ coated\ substrate\ mass - clean\ coated\ substrate\ mass}{surface\ area\ of\ coated\ substrate} \quad (2)$$

The values show a significant decrease of dust deposition rate (56% reduction) on the coated sample compared with the uncoated glass samples.

Table 2 shows that the measured film thickness of one process cycle of coating is 76.2 nm. This value is in good agreement with the film thickness obtained by Xi et al.[8] who first proposed this technique for self-cleaning applications. The obtained thickness for one process cycle in Xi et al. work was reported to be 70 ± 4 nm[8]. The process cycle was defined as dipping of the substrate in the colloidal solution for a soaking period of 2 h followed by a subsequent heating at 400 °C for another 2 h. If another cycle was required to form another coating layer, the coated sample would be dipped again in the colloidal solution for another 2 h and heated up again in air for 2 h, and so on, layer by layer. As mentioned earlier, this technique is simple, cost effective, straight forward and does not involve any sophisticated experimental equipment. As a matter of fact, Xi et al.[8] reported that they could successfully scale up the coating films on larger areas (125 mm × 125 cm × 2 mm) due to the simplicity of this coating method.

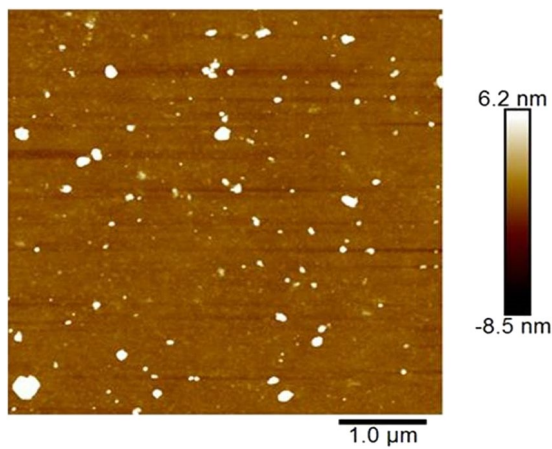

**Figure 5.** Top-view AFM image of TiO$_2$ thin film on glass substrate.

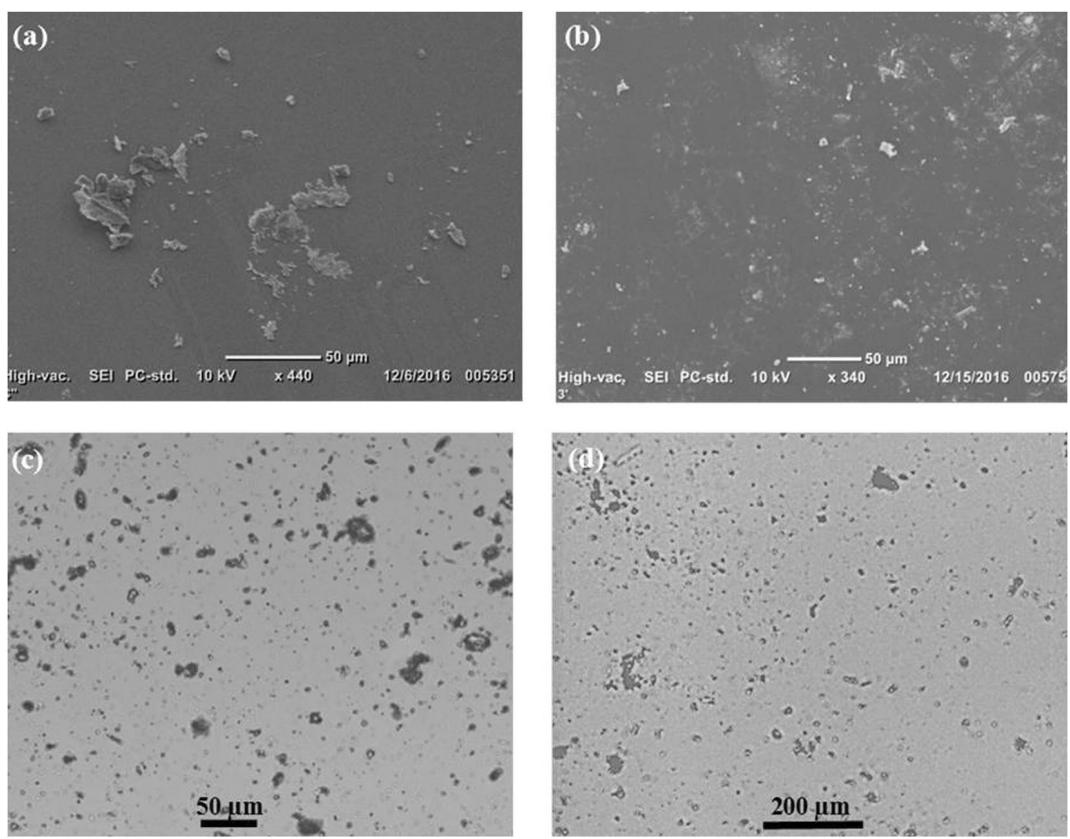

**Figure 6.** SEM images showing the top surface of the (**a**) uncoated and (**b**) coated glass samples, respectively. Optical images of glass substrate of the (**c**) uncoated and (**d**) coated with TiO$_2$ samples after soiling for 7 days.

The measured contact angles of the deposited films show medium hydrophilicity and the values are in total agreement with what was reported earlier for anatase TiO$_2$ thin films deposited on glass substrates as summarized in Table 3 [21–23]. The water spread is an important factor for self-cleaning as well as for manual cleaning with water as an outside cleaning source (Fig. 8).

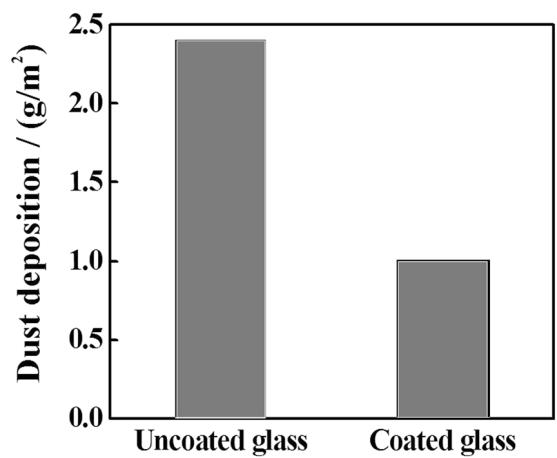

**Figure 7.** The average of dust deposition rate for seven days on uncoated versus coated samples with TiO$_2$.

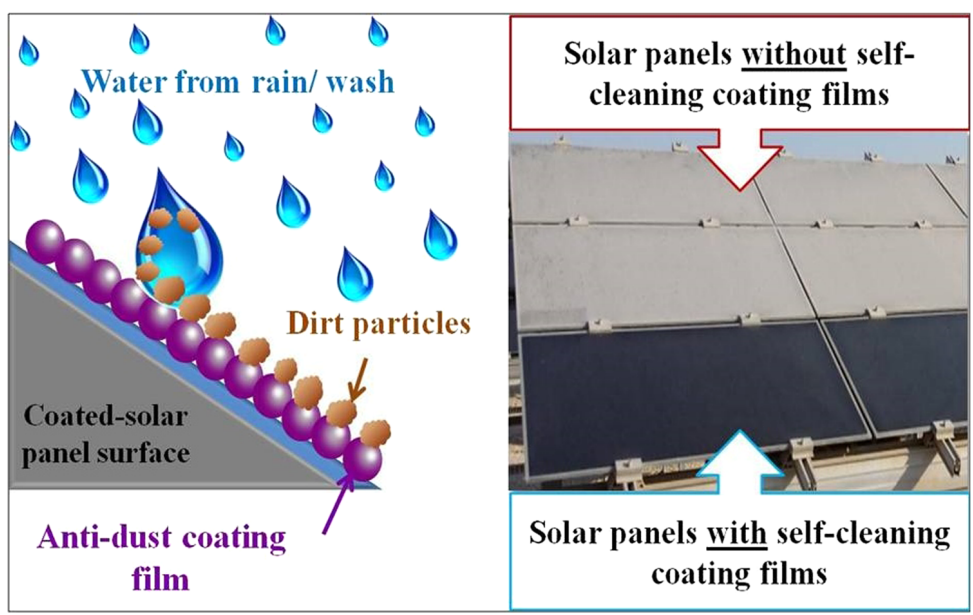

**Figure 8.** Schematic showing self-cleaning coating effect on solar panels in dusty regions.

| Coating | Film thickness (nm) | Refractive index |
|---|---|---|
| TiO$_2$ | 76.2 | 1.51 |

**Table 2.** Film thickness and refractive index of TiO$_2$ coating films.

| TiO$_2$ Film deposition method | Calcination temperature of TiO$_2$ films (°C) | Contact angle (°) | Reference |
|---|---|---|---|
| Adsorptive self-assembly | 400 | 43 | This work |
| Electrospinning | 500 | 43 | [21] |
| Liquid frame spray | 450 | 45 | [22] |
| Sol-gel dip coating | 450 | 48 | [23] |

**Table 3.** Summary of the measured contact angles of water on TiO$_2$ films produced by various methods.

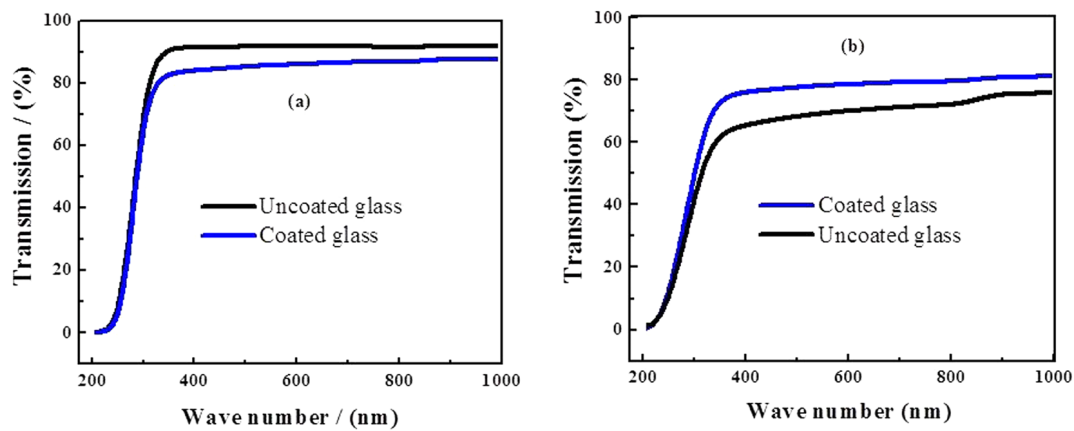

**Figure 9.** UV-Vis spectra of the samples (**a**) before soiling and (**b**) after soiling for 7 days.

It has been reported that the contact angles of the films deposited on glass substrates depend on the characteristics of the nanocolloids[10]. Salvaggio et al.[10] reported that increasing acid concentration from 0.1 to 0.5 N ($HNO_3$) during $TiO_2$ nanaoparticles synthesis lead to a decrease in the particle size from 31 to 20 nm. This decrease in the nanoparticle size triggered an increase in the surface roughness which in turn increased the contact angle from 44 to 60°. Besides, the increase in contact angle can be attributed to the formation of rough micro-nano-morphological pattern reduction thereby the surface energy of the particles[24].

In order to investigate the effect of sunlight exposure on the hydrophilic property of the coated samples, the contact angle was measured after the samples were left to soiling process for a week. The value of the contact angle is found to be around 41°. This slight reduction in the contact angle (which was measured to be 43° before light exposure) is rather expected[25, 26]. The effect of the sunlight has been reported earlier and was found to decrease the contact angle over time[25, 26]. On the other hand, when the samples were stored in dark conditions for several months, the contact angle was measured afterwards and found to increase up to 59°. This property of surface wettability conversion of $TiO_2$ has been extensively reported as well[27–29].

Literature survey has shown that contact angle of $TiO_2$, for either water and/or different liquid, decreases (surface becomes more hydrophilic) during UV irradiation with dependence on the intensity and time of irradiation. On the other hand, the hydrophilic surface can be reversed to a hydrophobic one (having higher contact angles) when it was kept in the dark conditions for a period of time[28, 29]. The mechanism of this hydrophilic conversion of titania surface was presented by Wang et al.[28] and Sakai et al.[30] who postulated that upon UV irradiation, the produced electrons and holes are trapped by the surface and $O^{2-}$ ions producing $Ti^{+3}$ and oxygen vacancies, respectively. This results in adsorption of water molecules at the defect sites forming a hydrophilic dominant surface.

To investigate the effect of coating on the optical transmission of the glass substrate, UV-Vis spectra was collected for the coated and uncoated glass surfaces before and after soiling for 7 days. The Spectra are shown in Fig. 9. The transmission of the clean glass had a maximum value of 91.75% in the range of values (90–92%) as reported in many studies[31, 32]. To investigate the effect of coating in mitigating the dust accumulation onto the coated substrates, the samples were left at The Solar Test Facility located in Doha/ The State of Qatar, for 7 days. It is shown in Fig. 7(b) that the coated sample shows an average reduction of only 6% in the visible region (400–800 nm) compared with clean uncoated surface. It has been reported in the literature that the increased surface roughness was responsible for the slight loss in optical transmission in the visible range[33].

The enhanced performance of the coated sample is due to several factors. First, the synthesized material is $TiO_2$ anatase which is well known for its high photocatalytic activity. The self-cleaning process using $TiO_2$ films involves two stages; first, a split of organic dirt via photocatalytic process in the presence of ultraviolet light. This stage is very critical as it reduces the presence of the sticky material that attracts the deposition of dust and other particulate material on the surfaces. The second factor is the hydrophilicity of $TiO_2$ films. Coating films of solar panels in desert and arid regions is preferred to be hydrophilic since further scheduled cleaning from time to time using water is essential to eliminate the dirt from the surface by the diffusal of water to the whole surface instead of getting together and hence rinsing the dust.

The low refractive index enhances transmittance which is one of the main factor for film coatings specifically applicable for solar panels[13, 32]. Table 2 shows the measured refractive index of 1.51. The antireflection properties which depend on the interference of the reflected light from air-coating and coating substrate interfaces were examined out. For an ideal homogeneous single coating which has a refractive index value between that of air and the substrate, the antireflection coating should satisfy the following conditions: the thickness of the coating should be $\lambda/4$, where $\lambda$ is the wavelength of the incident light; and $n_c = (n_a \times n_s)^{0.5}$, where $n_c$, $n_a$, and $n_s$ are the refractive indices of the coating, air, and substrate, respectively[34]. Taking into consideration that $n_a = 1$ for air, $n_s$ is 1.52 for glass substrate, $n_c$ must be 1.23 to achieve zero reflection. Since this value is lower than that of any homogeneous dielectric material, AR coatings always adopt 2- or 3-dimensional porous structures to meet the requirement for very low average refractive index[32, 34].

## Conclusions

Photocatalytic TiO$_2$ coatings are known for their excellent self-cleaning behavior and anti-reflection properties, where very thin water layer formed on the hydrophilic surface can easily wash-off the dirt particles. In the present work, we reported on a costly efficient and simple preparation method of the optically transparent, good wettability towards water and photocatalytic TiO$_2$ coatings on glass substrate for self-cleaning applications. TiO$_2$ nanocolloids particles were successfully synthesized via polyol reduction method and were subsequently used as coating onto borosilicate glass substrates via adsorptive self-assembly process. Our films were characterized by atomic force microscopy, scanning electron microscopy, ellipsometry and water contact angle measurements. The self-cleaning capability of the films was investigated after several days of dust accumulation by optical microscopy and UV-Vis spectroscopy. Our results showed that the coated glass samples have transmission loss of less than 6% along with 56% reduction in dust deposition rate compared with bare uncoated glass, paving the way to their application on PV panels, light-weight window and/or door polycarbonates for excellent self-cleaning applications.

### Acknowledgements
We acknowledge financial support from Qatar Environment and Energy Research Institute (QEERI), at Hamad Bin Khalifa University-Qatar Foundation (Doha, Qatar). Authors would like to thank Dr. Tarik Rhadfi for the synthesis of the nanoparticles, Dr. Said Mansour (Research Director of the Characterization and Imaging Group at QEERI) for performing TEM of the nanocolloids, and the office of research at Texas A&M University, Doha, for their fruitful collaboration.

### Author Contributions
R.I. helped with $TiO_2$ nanoparticle synthesis, performed the coating of the samples, optical microscopy, contact angle measurement, UV-Vis, analysis of data obtained and writing the article. A.S. performed ellipsometry measurements for film thickness and refractive index, W.S. obtained SEM images, D.J. performed AFM for surface topography and roughness of the films, W.Y. did XRD of the nanoparticles and deposited films, A.A obtained raw materials for the project, B.A. assisted with data analysis and writing the article.

### Additional Information
**Competing Interests:** The authors declare that they have no competing interests.

**Publisher's note:** Springer Nature remains neutral with regard to jurisdictional claims in published maps and institutional affiliations.